\documentclass[article, twocolumn, 
  amsmath,amssymb,
  longbibliography,
  ]{revtex4-1}
\usepackage{graphicx,color}
\usepackage{amsmath}
\usepackage{natbib}
\usepackage{epsfig}
\begin{document}

\title{Note: Gas-liquid crossover in the Lennard-Jones system}

\author{S. A. Khrapak}\email{Sergey.Khrapak@gmx.de}
\affiliation{Joint Institute for High Temperatures, Russian Academy of Sciences, 125412 Moscow, Russia}

\begin{abstract}
It is demonstrated that the crossover between gas- and liquid-like regions on the phase diagram of the Lennard-Jones system occurs at a fixed value of the density divided by its value at the freezing point, $\rho/\rho_{\rm fr}\simeq 0.35$. This definition is consistent with other definitions proposed recently. As a result a very simple practical expression for the gas-to-liquid crossover line emerges.  
\end{abstract}

\date{\today}
\maketitle


In recent years, the possibility to define a demarcation line between liquid-like and gas-like behaviors of supercritical fluids has been a topic of major interest~\cite{GorelliPRL2006,Simeoni2010,McMillanNatPhys2010,BrazhkinJPCB2011,BrazhkinPRE2012,BrazhkinUFN2012,
BrazhkinPRL2013,GorelliSciRep2013,YangPRE2015,
BrykJPCL2017,BrazhkinJPCB2018,BrykJPCB2018,BellJCP2020,
ProctorJPCL2019,PloetzJPCB2019,BanutiJSupFluids2020,HaJPCL2020,
MaximNatCom2019,SunPRL2020,KhrapakPRE04_2021,BellJPCL2021,
KhrapakPRE10_2021,CockrellPRE2021,CockrellPhysRep2021}. This problem is not just a matter of curiosity. The structural and dynamical properties are very different in these regimes and quite different approaches are required for their description. For example, transport properties in low density (gas-like) fluid regime are dominated by interatomic collisions and the Chapman-Enskog theory is appropriate~\cite{ChapmanBook,LifshitzKinetics}. Atoms move along straight trajectories between collisions, the collision properties being determined by the pairwise interaction potential. In dense fluids (liquid-like) regime the transport properties are dominated by collective interactions between the atoms. As a result, individual atoms spend a substantial amount of time in local minima of the multidimensional potential energy surface (a phenomenon often referred to as ``caging''). These local minima migrate with time to allow liquids to flow, but on considerably slower time scales. It is of great fundamental and practical interest to understand where the crossover between the gas-like and liquid-like regimes takes place. {This problem is also related to a long-standing debate about the nature of the supercritical fluid and a more general question ``What is liquid?''~\cite{BarkerRMP1976,Sengers1979,Woodcock2017}.}               

Multiple definitions for the gas-to-liquid crossover have been proposed in recent years~\cite{BrazhkinPRE2012,BrazhkinUFN2012,BellJCP2020,KhrapakPRE04_2021}. Not all of these definitions were consistent or universal, which generated a significant amount of debate. From the isomorph theory perspective~\cite{DyreJPCB2014}, it seems not unreasonable to assume that a demarcation line between gas-like and fluid-like regimes should itself be an approximate isomorph. As such, it should be characterized by a quasi-universal value of the excess entropy (actual entropy minus entropy of an ideal gas at the same density and temperature). {As we shall see, the excess entropy is indeed approximately constant along the crosssover line, although it passes through a near-critical region, where the Pearson correlation coefficient (which should be greater than $\simeq 0.9$ for Roskilde-simple systems) drops considerably~\cite{DyreJPCB2014,BellJPCB2019}.} Properly reduced structural and dynamical properties should also be quasi-invariant. This point of view has been elaborated in recent works~\cite{BellJCP2020,BellJPCL2021}. One of the possible definitions is based on the location of the minimum of the macroscopically scaled shear viscosity coefficient. This minimum corresponds to the crossover between the gas-like and liquid-like mechanisms of momentum transfer, which have quite different nature and hence different asymptotes. It has been observed that for several different model systems the location of the minimua in the macroscopically reduced shear viscosity coefficients occurs at approximately the same value of the excess entropy per particle ($s_{\rm ex}\simeq -2k_{\rm B}/3$)~\cite{BellJCP2020}, 
and the minimuma values themselves are also quasiuniversal~\cite{KhrapakPoF2022}. This is also where kinetic and potential contributions to the shear viscosity coefficients are equal to a good accuracy~\cite{BellJPCL2021}. Thus, the three considered definitions of a gas-like to liquid-like demarcation line (minima of reduced viscosity, constancy of excess entropy, and equality of kinetic and potential contributions to viscosity) nearly coincide with each other~\cite{BellJPCL2021}.            

Independently, it has been recently observed that in various simple fluids (such as Lennard-Jones, Coulomb, Yukawa and hard-sphere fluids) there exist two clear asymptotes for the product $D\eta(\Delta/k_{\rm B}T)$, where $D$ is the self-diffusion coefficient, $\eta$ is the shear viscosity coefficient, $\Delta=\rho^{-1/3}$ is the mean interatomic separation, $\rho$ is the atomic density, and $T$ is the temperature~\cite{KhrapakPRE10_2021}. In dense fluids near the freezing point this product approaches a slightly system-dependent constant value. This is the regime where the Stokes-Einstein relation without the hydrodynamic diameter is satisfied (in fact, interatomic separation plays the role of hydrodynamic diameter)~\cite{CostigliolaJCP2019,KhrapakMolPhys2019,KhrapakPRE10_2021}.  
Far away from the freezing point, in a rarefied fluid, this product decreases with increasing density. The intersection of the two asymptotes has been suggested as a convenient practical condition for the crossover between the gas-like and liquid-like regions on the phase diagram~\cite{KhrapakPRE04_2021,KhrapakPRE10_2021}. 
For the systems considered, intersection is characterised by very close values of the reduced excess entropy, $s_{\rm ex}/k_{\rm B}=-0.9 \pm 0.1$~\cite{KhrapakPRE10_2021}. {Similar value was obtained earlier in Ref.~\cite{BellJPCL2021}, where it was also pointed out that this is nearly the critical point entropy for simple fluids exhibiting a critical point.}

Let us now focus on the Lennard-Jones (LJ) system, a very important model in condensed matter studies. This model combines relative simplicity with adequate semi-quantitative approximation of interatomic interactions in real substances (such as condensed noble gases). The LJ potential is
\begin{equation}\label{LJ}
\phi(r)=4\epsilon\left[\left(\frac{\sigma}{r}\right)^{12}-\left(\frac{\sigma}{r}\right)^{6}\right], 
\end{equation}
where  $\epsilon$ and $\sigma$ are the energy and length scales (LJ units), respectively. The density and temperature expressed in LJ units are $\rho_*=\rho\sigma^3$ and $T_*=k_{\rm B}T/\epsilon$. 

It has been recently demonstrated that properly reduced transport coefficients of LJ fluids along isotherms exhibit quasi-universal scaling on the density divided by its value at the freezing point~\cite{KhrapakPRE04_2021}. In particular, transport coefficients such as self-diffusion, shear viscosity, and thermal conductivity, expressed in Rosenfeld's units~\cite{RosenfeldJPCM1999} remain approximately constant for fixed ratios $\rho/\rho_{\rm fr}$, where $\rho_{\rm fr}$ is the atomic number density at the freezing point (at a given temperature).  Combined with the excess entropy scaling of the transport coefficients~\cite{RosenfeldPRA1977,RosenfeldJPCM1999,DyreJCP2018,BellJPCB2019} this implies that the lines of constant ratio of $\rho/\rho_{\rm fr}$ are likely characterized by fixed values of excess entropy and thus are isomorphs. The minima of the reduced shear viscosity and thermal conductivity coefficients along isotherms correspond to $\rho/\rho_{\rm fr}\simeq 0.2 - 0.3$~\cite{KhrapakPRE04_2021}. The asymptotes of the product $D\eta(\Delta/k_{\rm B}T)$ are intersecting at $\rho/\rho_{\rm fr}\simeq 0.35$ and this can be considered as an appropriate definition of the crossover between the gas-like and the liquid-like regimes~\cite{KhrapakPRE04_2021,KhrapakPRE10_2021}. The Stokes-Einstein relation does not apply immediately after the crossover; even higher densities are required, such that $\rho/\rho_{\rm fr}\gtrsim 0.6$ for the LJ fluid~\cite{KhrapakPRE10_2021}. 

\begin{figure}
\includegraphics[width=7.2cm]{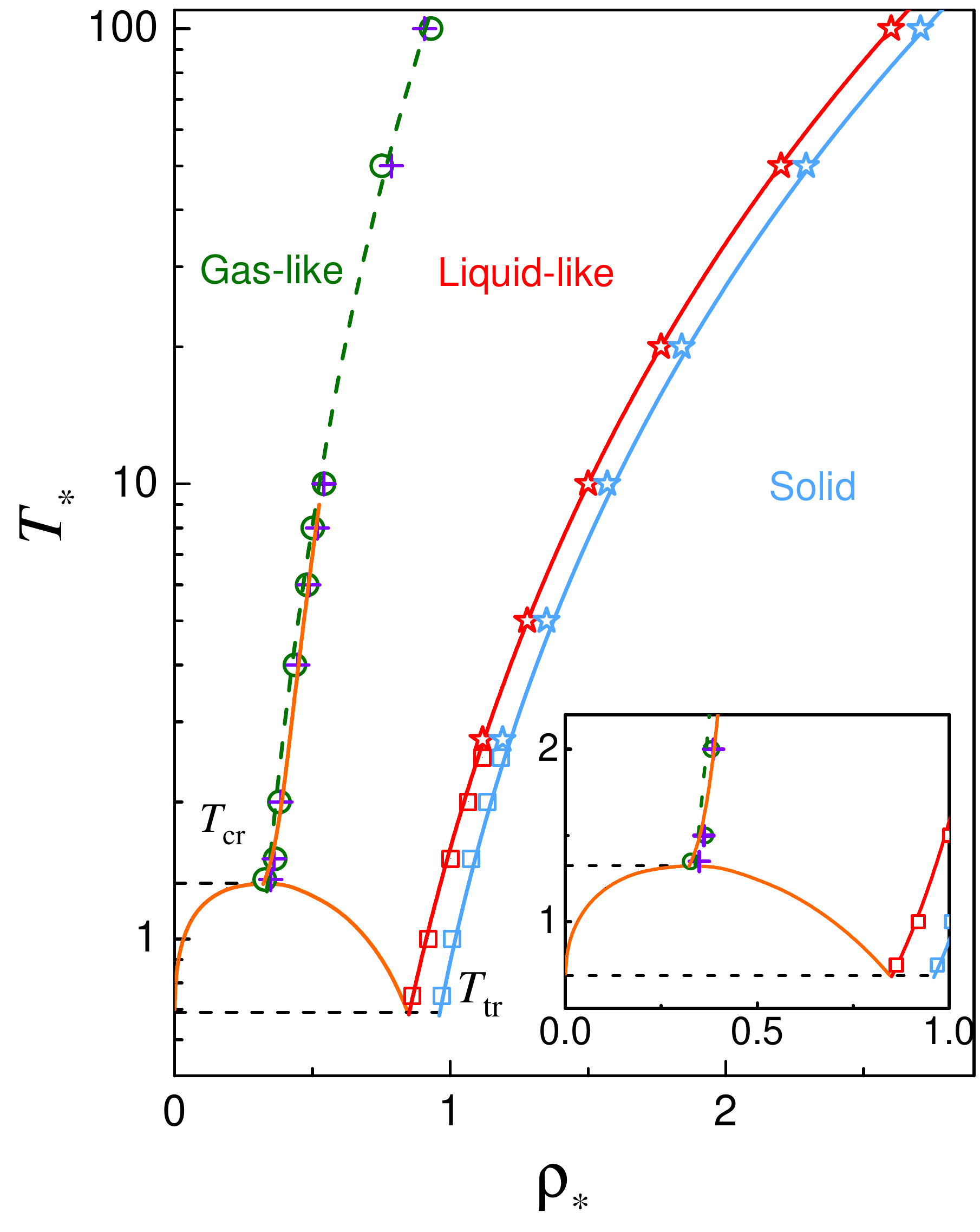}
\caption{(Color online) Phase diagram of the LJ system on the density-temperature plane. The squares and stars correspond to the fluid-solid coexistence boundaries as tabulated in Refs.~\cite{SousaJCP2012} and \cite{HansenPRA1970}, respectively; the corresponding curves are simple fits from Ref.~\cite{KhrapakJCP2011_2}. The liquid-vapour boundary is plotted using the formulas provided in Ref.~\cite{HeyesJCP2019}. 
The circles correspond to the location of minima of kinematic viscosity~\cite{BellJPCL2021}. The crosses denote the points where the contribution to viscosity due to atomic translation is exactly equal to that due to interatomic interactions~\cite{BellJPCL2021}. The dashed curve is plotted using Eq.~(\ref{Demarcation}) obtained from the constant density ratio $\rho/\rho_{\rm fr}=0.35$. A solid line emanating from the critical point corresponds to constant reduced excess entropy of $s_{\rm ex}=-0.9k_{\rm B}$, as evaluated from Thol's {\it et al}. equation of state~\cite{Thol2016}. {Inset shows a low temperature portion of the phase diagram with the focus on near-critical region.} All considered definitions of the crossover between the gas-like and liquid-like regimes of atomic dynamics are in excellent agreement.}
\label{Fig1}
\end{figure}

This hypothesis is verified in Fig.~\ref{Fig1}. It shows the phase diagram of the LJ system in the ($\rho_*$, $T_*$) plane. The fluid-solid coexistence data points are taken from Refs.~\cite{SousaJCP2012,HansenPRA1970}. The curves are the fits of the form  $T_*^{\rm L,S}= {\mathcal A}^{\rm L,S}\rho_*^4-{\mathcal B}^{\rm L,S}\rho_*^2$ (superscripts L and S correspond to liquid and solid, respectively ). This shape of the fluid-solid coexistence of LJ systems with constant (or very weakly $\rho_*$-dependent) constants ${\mathcal A}$ and ${\mathcal B}$ is a very robust result reproduced in a number of various theories and approximations~\cite{PedersenNatCom2016,RosenfeldMolPhys1976,KhrapakJCP2011_2,
HeyesPSS2015,KhrapakAIPAdv2016,CostigliolaPCCP2016,KhrapakPRR2020,
HeyesPRE2021}. Here we take constant values of ${\mathcal A}=2.29$ and ${\mathcal B}=0.71$ at freezing proposed in Ref.~\cite{KhrapakJCP2011_2}.  The liquid-vapour boundary is plotted using the formulas provided in Ref.~\cite{HeyesJCP2019}. The reduced triple point and critical temperatures are $T_*^{\rm tr}\simeq 0.694$~\cite{SousaJCP2012} and $T_*^{\rm cr}\simeq 1.326$~\cite{HeyesJCP2019}, respectively. Additional symbols appearing in the supercritical region are: The circles correspond to the location of minima of kinematic viscosity~\cite{BellJPCL2021}; the crosses denote the points where the contribution to viscosity due to kinetic and potential contributions are equal~\cite{BellJPCL2021}; the solid curve emanating from vary nearly the critical point and terminating at $T_*=9$ corresponds to constant excess entropy of $s_{\rm ex}=-0.9k_{\rm B}$ as calculated from the equation of state proposed by Thol {\it et al}.~\cite{Thol2016} in the domain of its applicability. The dashed curve corresponds to the condition $\rho/\rho_{\rm fr}=0.35$. In view of the relation between the freezing temperature and density a particularly simple expression for the crossover line emerges:
\begin{equation}\label{Demarcation}
T_*=2.29 (\rho_*/0.35)^4-0.71(\rho_*/0.35)^2.
\end{equation}          
Excellent agreement with other definitions of the gas-to-liquid crossover are observed from the critical point temperature up to the highest temperatures investigated ($T_*=100$). While different definitions agree, the definition based on the freezing density is generally more practical, because freezing density is usually relatively well known (as compared to the exact location of the minima in reduced transport coefficients, or lines of constant excess entropy).   

We observe in Fig.~\ref{Fig1} that the crossover line originates in the vicinity of the critical point. While this turns out to be quite convenient, this by no means represents a necessary requirement. The crossover occurs also in systems with purely repulsive interactions, where liquid-gas phase transition and the critical point are absent at all~\cite{BrazhkinPRL2013,YangPRE2015}. The crossover line would appear nearly parallel to the freezing line in the log-log plot. This relationship was pointed out and explained in Ref.~\cite{YangPRE2015}.  

To conclude, the crossover between the gas-like and liquid-like regions on the phase diagram of the Lennard-Jones system corresponds to fixed values of the density divided by its value at the freezing point and of the reduced excess entropy. The value $\rho/\rho_{\rm fr}=0.35$ and $s_{\rm ex}\simeq -0.9k_{\rm B}$, suggested previously from the analysis of the applicability of the Stokes-Einstein relation, provides very good agreement with other definitions (based on the minima of kinematic viscosity and equal contribution of the kinetic and potential terms to the viscosity coefficient). A very simple explicit expression for the demarcation line is derived. The obtained results can be useful in identifying gas-to-liquid crossover in liquified noble gases. Similar arguments are likely applicable to other simple systems beyond LJ. While particular values of $\rho/\rho_{\rm fr}$ may be different, the values of $s_{\rm ex}$ are expected to be approximately the same.     

I would like to thank V. Brazhkin for useful comments on the manuscript.

\bibliography{SE_Ref}

\end{document}